\documentclass[journal]{IEEEtran}

\usepackage{cite}
\usepackage{xcolor}

\ifCLASSINFOpdf
\usepackage[pdftex]{graphicx}
\graphicspath{{../figures/}}
\DeclareGraphicsExtensions{.pdf,.jpeg,.png}
\else
\fi

\usepackage{amsmath}
\usepackage{amssymb}

\usepackage[ruled]{algorithm2e}
\usepackage{algpseudocode}

\usepackage{array}

\usepackage{textcomp}
\usepackage{makecell}
\usepackage{hhline}

\usepackage{multicol}
\usepackage{multirow}
\usepackage{booktabs}

\usepackage{pifont}
\newcommand{\xmark}{\ding{55}}%

\newcommand{\norm}[1]{\left\lVert#1\right\rVert_2}
\newcommand{\vb}{\boldsymbol}

\newcommand{\mb}[1]{{\mathbf{#1}}}

\newcommand{\thickhline}{
    \noalign {\ifnum 0=`}\fi \hrule height 1pt
    \futurelet \reserved@a \@xhline
}

\SetKw{Continue}{continue}

\begin{document}

\title{Urban Road Safety Prediction: A Satellite Navigation Perspective}

\author{Halim Lee,
        Jiwon Seo,~\IEEEmembership{Member,~IEEE,}
        and Zaher M. Kassas,~\IEEEmembership{Senior Member,~IEEE,}
        \\
\thanks{H. Lee and J. Seo are with the School of Integrated Technology, Yonsei University, Incheon 21983, Republic of Korea,
E-mail: {\tt\small halim.lee@yonsei.ac.kr, jiwon.seo@yonsei.ac.kr}

Zaher M. Kassas is with Department of Mechanical and Aerospace Engineering, University of California, Irvine, CA 92697, USA,
E-mail: {\tt\small zkassas@ieee.org}. \textit{Corresponding authors: J. Seo and Z. Kassas.}
}}

\markboth{IEEE Intelligent Transportation Systems Magazine}%
{Urban Road Safety Prediction}

\maketitle

\begin{abstract}
Predicting the safety of urban roads for navigation via global navigation satellite systems (GNSS) signals is considered. To ensure safe driving of automated vehicles, the vehicle must plan its trajectory to avoid navigating on unsafe roads (e.g., icy conditions, construction zones, narrow streets, etc.). Such information can be derived from the roads' physical properties, vehicle's capabilities, and weather conditions. From a GNSS-based navigation perspective, the reliability of GNSS signals in different locales, which is heavily dependent on the road layout within the surrounding environment, is crucial to ensure safe automated driving. An urban road environment surrounded by tall objects can significantly degrade the accuracy and availability of GNSS signals. This article proposes an approach to predict the reliability of GNSS-based navigation to ensure safe urban navigation. Satellite navigation reliability at a given location and time on a road is determined based on the probabilistic position error bound of the vehicle-mounted GNSS receiver.
A metric for GNSS reliability for ground vehicles is suggested, and a method to predict the conservative probabilistic error bound of the GNSS navigation solution is proposed.
A satellite navigation reliability map is generated for various navigation applications. As a case study, the reliability map is used in the proposed optimization problem formulation for automated ground vehicle safety-constrained path planning.

\end{abstract}


\section{Road Information for Navigation Safety}
For safe and reliable control of automated ground vehicles, various road information need to be estimated. Road information typically include road surface conditions such as dryness, wetness, and ice as well as road shapes such as curvature, bank angle, and slope angle. Satellite-based navigation reliability should also be considered as an important road information, because automated vehicles use various navigation sensors that are dependent on positioning, navigation, and timing (PNT) from global navigation satellite systems (GNSS). In particular, reliable and accurate GNSS-derived position is crucial for short-range driving control and long-range navigation and path planning, while timing is crucial for on-board sensor fusion, cooperative planning and control, and information exchange with other vehicles and the infrastructure. The reliability and accuracy of received GNSS signals is heavily dependent on the road layout within the surrounding environment.

An automated vehicle usually relies on GNSS, such as the Global Positioning System (GPS) of the U.S., GLONASS of Russia, Galileo of Europe, and Beidou of China, to obtain its \textit{absolute} position on Earth. Although other sensors such as vision \cite{Kim14:Multi-UAV,Li21:Visual}, radar \cite{Shin17:Autonomous,Feng19:Lane}, lidar \cite{Joshi18:Generation,Peng19:Modeling}, and ultrasonic \cite{Rhee19:Low-cost} sensors or sensor network \cite{Wang19:An,Wang22:A} can measure relative distances to nearby objects, GNSS receivers are the primary sensing modality for determining a vehicle's absolute position. This absolute position information is crucial, especially for initializing urban navigation processes using other sensors.
For example, given a GNSS position solution, one can narrow down the search space in digital maps, which are used with three-dimensional (3D) point clouds from a scanning lidar, to estimate in real-time the vehicle's position and heading to a lane-level accuracy to avoid collisions \cite{Montemerlo08:Junior}. In addition, when integrated with vision simultaneous localization and mapping (vSLAM) \cite{Li21:Visual}, GNSS can mitigate the accumulative positioning error. Furthermore, GNSS measurements can be used to fix the drift of inertial measurement units (IMUs) for determining the vehicle's linear and angular motion \cite{Atia17:Low-Cost,Kassas17:Robust}.


\begin{table*}[t!]
\centering
\caption{Comparison of GNSS Reliability Prediction Methods.}
\label{tab:Comparison}
\vspace{-4mm}
\begin{center}
{\renewcommand{\arraystretch}{1.4}
 \begin{tabular}[c]{>{\centering\arraybackslash}m{1.5cm}|>{\centering\arraybackslash}m{5.1cm}>{\centering\arraybackslash}m{4.5cm} >{\centering\arraybackslash}m{5.1cm}}
 \Xhline{2\arrayrulewidth}
    \thead{Method} & \thead{Metric for GNSS reliability} & \thead{Considered obstacles}
       & \thead{Verification method}\\
\Xhline{1.5\arrayrulewidth}
{Shetty and Gao \cite{Shetty20:Predicting}} &
State uncertainty bound (3$\sigma$) that encloses the uncertain future state distributions & Buildings in virtual urban environment & Simulations only \\
 \hline
 {Zhang and Hsu \cite{Zhang19:New}} &
GPS positioning error & Real-world buildings without the consideration of driving lanes & Experiments (mean of the measured and predicted positioning errors differed by a maximum of 17.7 m) \\
 \hline
 {Maaref and Kassas \cite{Maaref20:Optimal}} &
GPS HPL without the consideration of measurement faults & Not considered (all GPS signals were assumed to be direct LOS) & Experiments (no performance comparison between the predicted and measured HPLs was reported) \\
\hline
 {Lee \textit{et al.} \cite{Lee20:Integrity}} &
GPS HPL with the consideration of multiple measurement faults (FDE was not performed) & Real-world buildings without the consideration of driving lanes & Simulations only \\
  \hline
 {Proposed} &
Conservative multi-constellation GNSS HPL with the consideration of multiple measurement faults (FDE was performed) & Real-world buildings and surrounding vehicles with the consideration of driving lanes &  Experiments (conservatively predicted HPL bounded the measured HPL in 100\% of the time) \\
 \Xhline{2\arrayrulewidth}
\end{tabular}}
\end{center}
\end{table*}

GNSS and differential correction stations alone can provide centimeter-level positioning accuracy if the signal reception environment and solar activity are favorable \cite{Jackson18:Performance}. Urban canyons impose harsh signal reception conditions \cite{Costa11:Simulation}. Tall buildings, trees, and nearby vehicles frequently block GNSS signals. Non-line-of-sight (NLOS) reception of GNSS signals without the reception of line-of-sight (LOS) signals, i.e., \textit{NLOS-only condition}, which occasionally occurs on urban roads, can cause arbitrarily large position errors.
In addition, the accuracy of pseudoranges (i.e., measured distances between the user's receiver and GNSS satellites without compensating for the receiver's clock bias and atmospheric delays) is degraded in an urban environment where LOS and NLOS signals are simultaneously received, i.e., \textit{LOS+NLOS condition}.
Therefore, it is important to predict the reliability of GNSS signals on urban roads to ensure safe operation of automated ground vehicles.

Various studies have utilized 3D building models with or without ray tracing to overcome the unfavorable GNSS signal reception conditions in urban environments \cite{Betaille13:A,Wang13:GNSS,Zhang19:New,Strandjord20:Improved,Shetty20:Predicting}. Power matching \cite{Saab06:Power}, shadow matching \cite{Wang13:GNSS}, specular matching \cite{Strandjord20:Improved}, and urban trench modeling \cite{Betaille13:A} were developed to decrease the positioning error by predicting the NLOS conditions of GNSS satellites using a 3D building map. In \cite{Shetty20:Predicting} and \cite{Zhang19:New}, 3D building models along with ray-tracing techniques were utilized to predict pseudoranges at a given location in an urban multipath environment. The future state uncertainty \cite{Shetty20:Predicting} and predicted positioning error \cite{Zhang19:New} were then calculated based on the predicted pseudoranges. However, while GNSS signal blockage due to buildings was considered, blockage due to other objects (e.g., trees and nearby vehicles) was not considered, nor did the predicted positioning error consider the detection and exclusion of possible faulty satellite signals or the probabilistic error bound of the predicted position solution.


The probabilistic error bound of the GNSS position solution, which is referred to as the protection level (PL), as well as the concept of navigation integrity have been actively studied for safety-critical applications, such as aviation \cite{Walter08:Worldwide,Lee17:Monitoring}. In \cite{Maaref20:Optimal}, a receiver autonomous integrity monitoring (RAIM) algorithm was developed to predict the horizontal position error bound (i.e., horizontal PL (HPL)), as a measure of satellite navigation reliability for ground vehicles. However, this algorithm did not perform fault detection and exclusion (FDE), nor it considered multiple signal faults, which are expected in urban environments. Furthermore, urban NLOS-only and LOS+NLOS conditions were not considered, and it was assumed that all GPS signals were received by direct LOS.

To overcome these limitations, a multiple hypothesis solution separation (MHSS) RAIM method was applied in \cite{Lee20:Integrity}, which considered multiple signal faults to predict the HPL. However, FDE was still not performed, and the performance of the proposed method was not validated experimentally. Upon attempting to validate this method experimentally, it was discovered that the method did not accurately predict the HPL. This was due to the complexity of predicting the multipath environment sufficiently accurately and due to signal blockage owing to tall objects other than buildings.
As presented in Table \ref{tab:Comparison}, the method proposed in the current study addresses the aforementioned issues.


The contributions of this study are summarized as follows:
\begin{list}{$\bullet$}{\leftmargin=1em \itemindent=0em}
    \item A conservatively predicted multi-constellation GNSS HPL after detecting and excluding multiple signal faults is suggested as a metric for GNSS reliability for ground vehicles.
    This metric considers more realistic urban GNSS signal environments than the other metrics in Table \ref{tab:Comparison}.
    \item A method to conservatively predict GNSS HPLs for ground vehicles is proposed. While performing ray-tracing simulations with 3D urban digital maps, possible driving lanes and surrounding vehicles were considered and the most conservative value was selected at each longitudinal location along the test roads.
    \item It was experimentally shown that the proposed metric (i.e., conservatively predicted HPL) successfully overbounded the HPL calculated using real pseudorange measurements during the field tests in two cities.
    \item An optimization problem formulation for safety-constrained path planning is proposed. Unlike the previous studies, the unavailability of GNSS signals and continuous GNSS signal outages are considered in the problem formulation. A specific implementation to solve this problem is also presented and experimentally demonstrated. The proposed method enables automated ground vehicles to select the path that ensures navigation safety.
\end{list}

The rest of this article is organized as follows.
Section \ref{Prediction_of_Satellite_Navigation_Reliability_on_Urban_Roads} formulates the proposed approach to predict GNSS satellite signal reliability on urban roads along with how to conservatively predict the HPL. It also evaluates the conservatively predicted HPL versus experimentally measured HPL by a ground vehicle.
Section \ref{Application_Case_Study_Safety_Constrained_Path_Planning} presents an application case study of the proposed approach in the context of safety-constrained path planning.
An optimization problem is formulated, solved, and experimentally demonstrated.
Section \ref{Conclusion} presents concluding remarks.

\section{Prediction of Satellite Navigation Reliability on Urban Roads}
\label{Prediction_of_Satellite_Navigation_Reliability_on_Urban_Roads}
A GNSS receiver estimates its 3D position and clock bias using pseudorange measurements from at least four GNSS satellites. Because a pseudorange is directly related to the signal travel time from the satellite to the user's receiver, which is measured by a receiver clock, various errors, such as satellite clock bias and ionospheric and tropospheric delay errors, contaminate the pseudorange measurement. These errors should be corrected for to bring the pseudorange closer to the true range. The receiver clock bias is treated as an additional unknown variable, which is obtained alongside the receiver position through a solution estimation process. This section presents various error sources for satellite navigation systems and introduces the proposed method to predict pseudoranges and conservative position error bounds as a measure of satellite navigation reliability on urban roads.

\subsection{Error Sources for Satellite Navigation}

The performance of GNSS-based navigation can be degraded by anomalous ionospheric behavior \cite{Seo14:Future,Sun21:Markov,Lee22:Optimal}, radio frequency interference \cite{Park21:Single,Kim22:First}, signal reflection or blockage \cite{Lesouple18:Multipath,Zhang21:Performance}, and poor geometric diversity of satellites in view \cite{Saab02:Map,Bresler16:GNSS}. In particular, signal reflection or blockage due to buildings and other tall objects is a significant error source for ground vehicle navigation in urban canyons. When $N$ GNSS satellites are in view, the $n$-th pseudorange measurement in an urban environment at time-step $t$, after satellite clock bias corrections, can be modeled as follows
\begin{eqnarray}
\nonumber \rho^{n}(t) &=& R_\mathrm{LOS}^{n}(t) + \rho_\mathrm{bias}^{n}(t) + \varepsilon^{n}(t)\\
\nonumber    &=& \norm{ \vb{r}_{u}(t) - \vb{r}^{n}(t) } + c \cdot \delta t_{u}(t) \\
    &\,& + \, I^{n}(t) + T^{n}(t)  + \rho_\mathrm{bias}^n(t) + \varepsilon^{n}(t),
\end{eqnarray}
where the descriptions of the symbols are given in Table \ref{tab:NotationsPseudorange}.

\begin{table}
\centering
\caption{Mathematical Notations Related to the Pseudorange Measurement Modeling in Urban Environments.}
\label{tab:NotationsPseudorange}
\vspace{-6mm}
\begin{center}
{\renewcommand{\arraystretch}{1.4}
 \begin{tabular}[cl]{>{\centering\arraybackslash}m{1.0cm}|>{\arraybackslash}m{7cm}}
 \Xhline{2\arrayrulewidth}
    \thead{Symbol} & \thead{Description} \\
\Xhline{1.5\arrayrulewidth}
{$\rho^{n}$} & {$n$-th pseudorange measurement in an urban environment after satellite clock bias corrections} \\
{$R_\mathrm{LOS}^{n}$} & {length of the LOS path between the user's receiver and $n$-th satellite including delays due to receiver's clock bias, ionosphere, and troposphere} \\
{$\rho_\mathrm{bias}^{n}$} & {Either (i) the bias due to an NLOS-only condition (i.e., $\rho_\mathrm{NLOS}^{n}$) which represents the extra travel distance of the NLOS signal compared with $R_\mathrm{LOS}^{n}$ [see Fig. \ref{fig:multipath_and_NLOS} (top)] or (ii) the bias due to an LOS+NLOS condition (i.e., $\rho_\mathrm{L+N}^{n}$) where both LOS and NLOS signals are received [see Fig. \ref{fig:multipath_and_NLOS} (bottom)]} \\
{$\rho_\mathrm{NLOS}^{n}$} & {Bias due to an NLOS-only condition} \\
{$\rho_\mathrm{L+N}^{n}$} & {Bias due to an LOS+NLOS condition} \\
{$\vb{r}_u$} & {Position vector of the user's receiver} \\
{$\vb{r}^{n}$} & {Position vector of the $n$-th satellite} \\
{$c$} & {Speed of light} \\
{$\delta t_u$} & {User's receiver clock bias} \\
{$I^n$} & {Ionospheric delay in the $n$-th pseudorange measurement} \\
{$T^n$} & {Tropospheric delay in the $n$-th pseudorange measurement} \\
{$\varepsilon^{n}$} & {Remaining errors (e.g., noise, unmodeled effects, etc.) in the $n$-th pseudorange measurement} \\
 \Xhline{2\arrayrulewidth}
\end{tabular}}
\end{center}
\end{table}

\begin{figure}
    \centering
    \includegraphics[width=5.7cm]{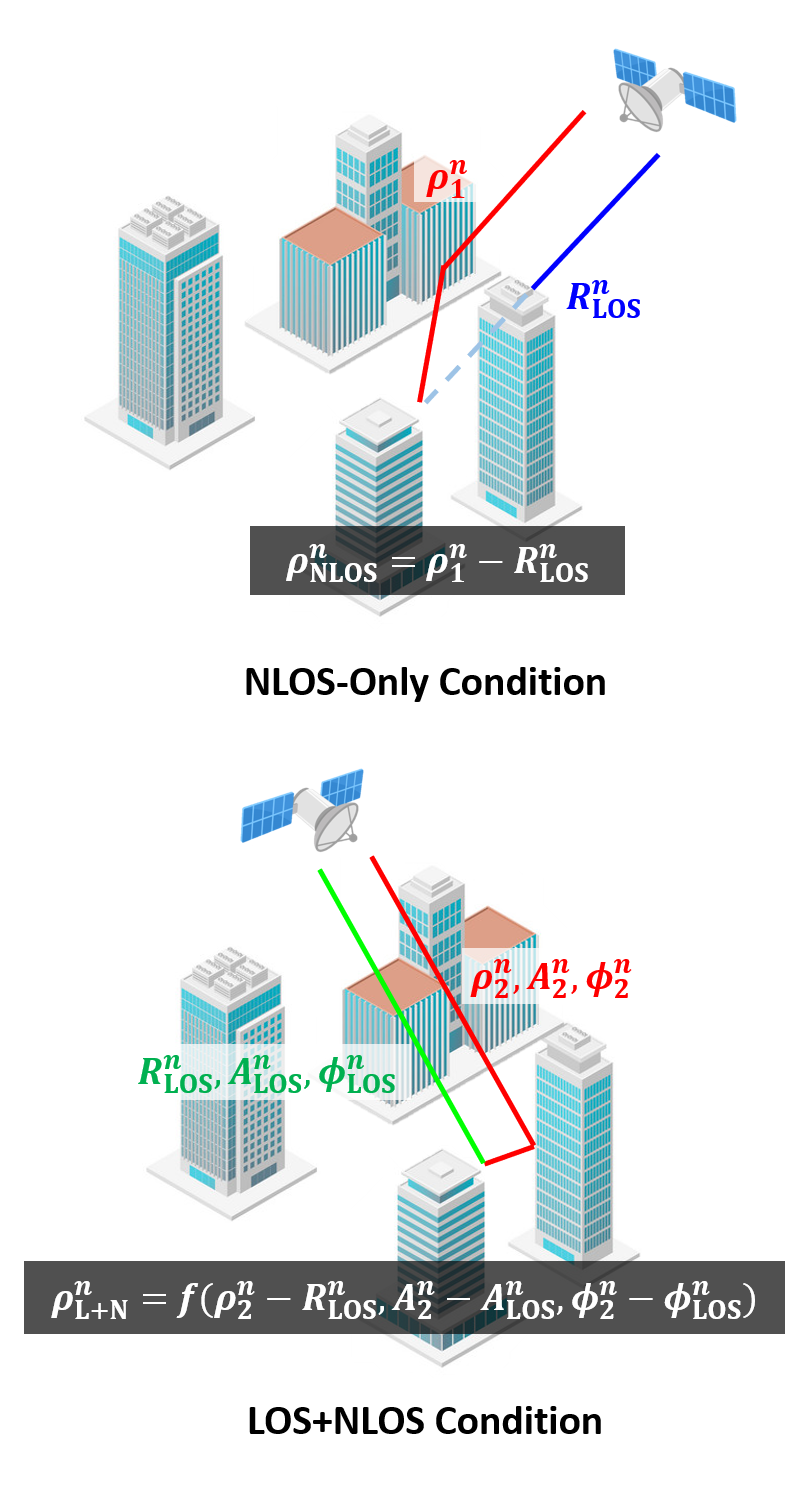}
    \caption{GNSS NLOS-only and LOS+NLOS conditions in an urban environment and corresponding pseudorange biases.}
    \label{fig:multipath_and_NLOS}
\end{figure}

Considerable common-mode errors can exist between a user and a nearby reference station, such as atmospheric delays and satellite ephemeris errors. These errors can be largely mitigated using differential GNSS (DGNSS). A DGNSS reference station broadcasts correction messages to nearby users, allowing the users to eliminate common-mode errors. However, site-specific errors caused by NLOS-only or LOS+NLOS signal reception cannot be mitigated using DGNSS.

Four GNSS signal reception conditions can occur in urban canyons: 1) LOS-only condition in which only the LOS signal is received, 2) NLOS-only condition in which only NLOS signals are received, 3) LOS+NLOS condition in which both the LOS and NLOS signals are received, and 4) no-signal condition in which the signal is completely blocked by an object. Fig. \ref{fig:multipath_and_NLOS} illustrates the difference between the NLOS-only and LOS+NLOS conditions. In the field of satellite navigation, the NLOS-only and LOS+NLOS conditions are treated differently as they cause different types of pseudorange errors. Moreover, simulation methods to predict these errors are different, as will be discussed next.

Under the NLOS-only condition, the NLOS-only bias term, which is $\rho_\mathrm{NLOS}^{n}$ in Fig. \ref{fig:multipath_and_NLOS} (top), reflects the extra travel distance (i.e., $\rho_1^n - R_\mathrm{LOS}^{n}$ where $\rho_1^n$ is the travel distance along the reflected path) due to signal reflection, which can be arbitrarily large. If this bias remains in the pseudorange measurement, it can cause a large unbounded positioning error. A typical way to predict $\rho_\mathrm{NLOS}^{n}$ at a given location is to calculate the difference between the lengths of the direct and reflected paths (i.e., LOS and NLOS paths) from a satellite to a receiver, which represents the extra travel distance. Ray-tracing simulation using 3D urban digital maps can be performed to estimate the length of the reflected path. The positions of the satellites at a given time for ray-tracing simulation are calculated based on the satellite broadcast almanac information. The complete blockage of the signal (i.e., no-signal condition) can also be predicted by ray-tracing simulation.

In an urban environment, the LOS+NLOS condition is more frequently observed than the NLOS-only condition. Unlike the NLOS-only bias term, the LOS+NLOS bias term, which is $\rho_\mathrm{L+N}^{n}$ in Fig. \ref{fig:multipath_and_NLOS}, is bounded. Reflected signals with a large delay compared with 1.5 chip width of the GNSS signal (e.g., 300 m width for GPS L1 C/A-code chip) do not cause any bias in the pseudorange measurements if the direct signal is also received and tracked \cite{Misra10:Global}. For the short-delay reflected signals (i.e., delay is less than 1.5 chips), $\rho_\mathrm{L+N}^{n}$ depends on the receiver's correlator design, and it is a function of the difference of travel distances (i.e., $\rho_{2}^{n} - R_\mathrm{LOS}^{n}$), received signal amplitudes (i.e., $A_{2}^{n} - A_\mathrm{LOS}^{n}$), and phases (i.e., $\phi_{2}^{n} - \phi_\mathrm{LOS}^{n}$) of reflected and direct signals, where $\left(\cdot\right)_2^n$ and $\left(\cdot\right)_\mathrm{LOS}^n$ represent the reflected and direct signals from the $n$-th satellite, respectively (see Fig. \ref{fig:multipath_and_NLOS} (bottom)).

The receiver used in the field experiments of this study, which will be explained in Section \ref{sec:FieldTest}, utilizes the \textit{a posteriori} multipath estimation (APME) method \cite{Sleewaegen01:Mitigating}; therefore, the multipath error envelop of the AMPE method was used to predict $\rho_\mathrm{L+N}^{n}$ in this study. The amplitudes and phases of the received reflected and direct signals were obtained through ray-tracing simulations.

\subsection{Probabilistic Error Bound and ARAIM}

The accuracy in the field of navigation usually refers to the 95th percentile value of the positioning error distribution \cite{Blanch07:Optimized}. However, when navigation safety is of concern, a considerably higher probability (e.g., 99.99999\% for the vertical guidance of aircraft) should be considered to obtain an error bound \cite{Walter08:Worldwide}. This error bound (i.e., PL) includes the true position of a user with a required high probability. If the PL is larger than the alert limit (AL) of a certain safety-critical operation (e.g., 35 m for the vertical guidance of an aircraft down to 200 ft above the runway), the position output from the navigation system is deemed unreliable because it is not guaranteed that the true position is within the AL with the required probability. In this case, the navigation system is declared unavailable and must not be used to ensure navigation safety (i.e., navigation integrity is guaranteed by a timely alert).

Among various methods and augmentation systems (e.g., ground based augmentation system (GBAS) \cite{Lee11:Ionospheric,Seo12:Targeted,Lee17:Real-time} and satellite based augmentation system (SBAS) \cite{Toledo07:High,Walter17:Improved}) to guarantee the integrity of satellite navigation systems, RAIM is often preferred because it requires no or minimal support from infrastructure. The basic idea of RAIM is to check the consistency between the position solutions obtained by subsets of pseudorange measurements. If all the subset solutions are almost identical, all the signals can be confirmed to be fault-free, and the position output of a receiver is deemed reliable.

\begin{figure}
    \centering
    \includegraphics[width=7.7cm]{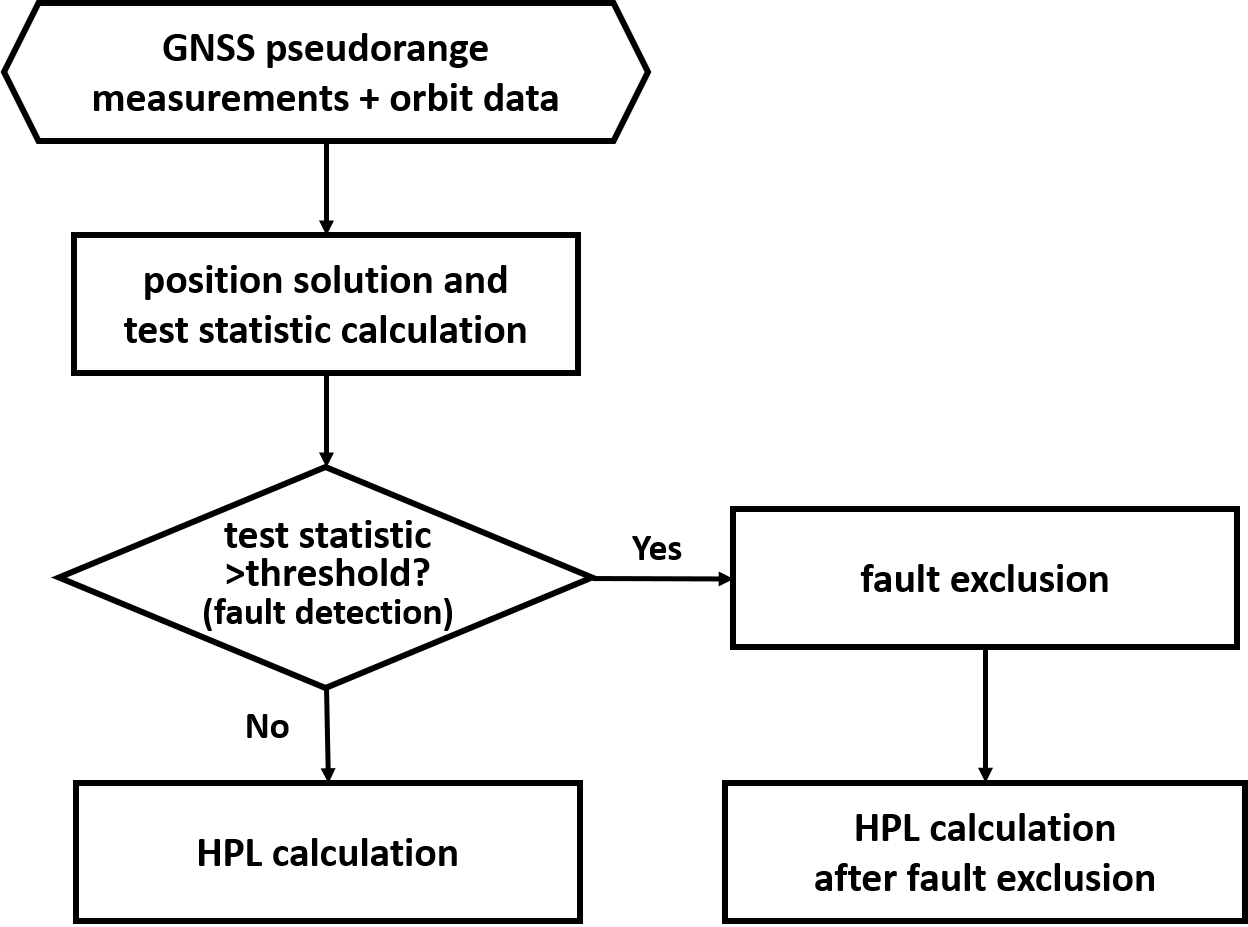}
    \caption{Flowchart of fault detection and exclusion (FDE) and horizontal protection level (HPL) calculation of a RAIM algorithm.}
    \label{fig:flowchart}
\end{figure}

Many RAIM algorithms have the functionality of FDE and PL calculations. FDE rejects faulty signals that cause erroneous position solutions through a consistency check using redundant measurements. A minimum of six pseudorange measurements are necessary to detect and exclude a single fault. PL is a probabilistic error bound of a position solution, and HPL is particularly relevant to ground vehicles. For aerial vehicles, the vertical PL (VPL) should be also considered \cite{Morales16:Opportunity,Maaref20:Aerial}. After performing FDE, the HPL can be calculated as shown in the flowchart in Fig. \ref{fig:flowchart}.

It should be noted that RAIM is suitable for real-time integrity monitoring of received GNSS signals; however, the focus of this study is not on guaranteeing real-time navigation integrity. Instead, a method is proposed to predict satellite navigation reliability at each location on urban roads before an automated vehicle arrives at that location in this study. The probabilistic position error bound (i.e., HPL) is used as a safety metric to represent the satellite navigation reliability. After the reliability is predicted and provided to the vehicle as part of the road information, the vehicle can detour the low-reliability region (i.e., high HPL region) or prepare its other navigation sensors to not utilize GNSS measurements when passing through the low-reliability region. 

For this purpose, ARAIM with a multiple hypothesis solution separation (MHSS) algorithm \cite{Blanch07:Optimized,Blanch15:Baseline} that can handle multiple faults and constellations is adopted in this study.
It is expected that a ground vehicle will experience multiple GNSS signal faults on urban roads. Currently, most GNSS receivers used by automated vehicles are capable of tracking multiple GNSS constellations (e.g., GPS and GLONASS were used in this study). By introducing multiple hypotheses of signal failures, ARAIM can detect and exclude multiple faults in multiple constellations and consider the possibility of further fault modes when calculating the HPL. Therefore, ARAIM among various RAIM algorithms \cite{Zhu18:GNSS} is appropriate for FDE based on the predicted pseudoranges and HPL prediction for automated ground vehicles in urban environments.

The MHSS-based FDE algorithm detects faulty signals using a solution separation threshold test. Solution separation is the difference between fault-free and fault-tolerant position solutions.
The receiver's state $\vb{x}$, which is $\vb{\hat{x}} + \Delta \vb{\hat{x}}$, can be estimated by the weighted least-squares estimator whose update equation is given by \cite{Blanch15:Baseline, Misra10:Global}
\begin{equation} \label{WLS}
    \Delta \vb{\hat{x}} = (\mb{G}^{\mathsf{T}} \mb{W} \mb{G})^{-1} \mb{G}^{\mathsf{T}} \mb{W} \Delta \vb{\rho},
\end{equation}
where the descriptions of the symbols are given in Table \ref{tab:NotationsARAIM}.
The fault-free position solution is estimated from the all-in-view satellites, whereas the fault-tolerant position solution assumes one or more possible faulty signals; thus, it is estimated from a subset of satellites. Then, the solution separation threshold test is expressed as \cite{Blanch15:Baseline}
\begin{equation} \label{SStest}
    \lvert \hat{x}_q^{(0)} - \hat{x}_q^{(k)} \rvert \leq
    T_{k,q},
\end{equation}
where the descriptions of the symbols are given in Table \ref{tab:NotationsARAIM}.
If the solution separation for any axis exceeds a certain threshold, signal faults are likely to exist, and exclusion of these faults should be attempted.

\begin{table}
\centering
\caption{Mathematical Notations Related to HPL Calculation.}
\label{tab:NotationsARAIM}
\vspace{-6mm}
\begin{center}
{\renewcommand{\arraystretch}{1.4}
 \begin{tabular}[cl]{>{\centering\arraybackslash}m{2.5cm}|>{\arraybackslash}m{5.5cm}}
 \Xhline{2\arrayrulewidth}
    \thead{Symbol} & \thead{Description} \\
\Xhline{1.5\arrayrulewidth}
{$\vb{x}$} & {State vector of the user's receiver, which is defined as $\left[\vb{r}_u^{\mathsf{T}}, c \delta t_u \right]^{\mathsf{T}}$} \\
{$\vb{\rho}$} & {Pseudorange measurement vector, which is defined as $\left[\rho^1, \ldots, \rho^N \right]^{\mathsf{T}}$} \\
{$\Delta \vb{\hat{x}}$} & {Difference between the receiver's state vector $\vb{x}$ and its estimate from the previous iteration $\vb{\hat{x}}$} \\
{$\Delta \vb{\rho}$} & {Difference between the pseudorange measurement vector $\vb{\rho}$ and the expected pseudorange vector $\vb{\hat{\rho}}$ based on the satellite positions and $\vb{\hat{x}}$} \\
{$\mb{G}$} & {Geometry matrix} \\
{$\mb{W}$} & {Weighting matrix, which is the inverse of a diagonal matrix whose diagonal elements are the measurement noise variances} \\
{$q$} & {$q = 1$ or $q = 2$ for the East or North axis of the horizontal plane, respectively} \\
{$\hat{x}_q^{(0)}$} & {Fault-free position solution for the $q$ axis estimated from the all-in-view satellites} \\
{$\hat{x}_q^{(k)}$} & {Fault-tolerant position solution for the $q$ axis and $k$-th fault mode} \\
{$T_{k,q}$} & {Solution separation threshold for the $q$ axis and $k$-th fault mode ($k=0$ represents the fault-free condition)} \\
{$H\!P\!L_q$} & {HPL for the $q$ axis} \\
{$\mb{Q}(\cdot)$} & {Tail probability function of the standard Gaussian distribution} \\
{$b_{q}^{(k)}$} & {Nominal bias of the position solution for the $q$ axis and $k$-th fault mode} \\
{$\sigma_{q}^{(k)}$} & {Standard deviation of the position solution for the $q$ axis and $k$-th fault mode} \\
{$N_\mathrm{fault \: modes}$} & {Total number of fault modes} \\
{$p_{\mathrm{fault},k}$} & {Probability that the $k$-th fault mode occurs} \\
{${P\!H\!M\!I}_\mathrm{HOR}$} & {Probability of hazardously misleading information for the horizontal component} \\
{${P\!H\!M\!I}_\mathrm{VERT}$} & {Probability of hazardously misleading information for the vertical component} \\
{$P_\mathrm{sat, not \: monitored}$} & {Probability that independent simultaneous satellite faults are not monitored} \\
{$P_\mathrm{const, not \: monitored}$} & {Probability that simultaneous constellation faults are not monitored} \\
 \Xhline{2\arrayrulewidth}
\end{tabular}}
\end{center}
\end{table}

If the solution separation threshold test passes without excluding any satellite signals, the HPL is computed as follows. In the MHSS-based HPL calculation method, HPL is obtained as a bound that includes all the HPLs corresponding to the fault-free and fault-tolerant position solutions. The HPL for the $q$ axis (i.e., $H\!P\!L_q$) is calculated as \cite{Blanch15:Baseline}
\begin{equation} \label{eq:HPL}
\begin{split}
    & 2\mb{Q}\left(\frac{{H\!P\!L}_{q} - b_{q}^{(0)}}{\sigma_{q}^{(0)}}\right)\\
    &+\sum_{k=1}^{N_\mathrm{fault \: modes}} p_{\mathrm{fault},k} \mb{Q} \left(\frac{H\!P\!L_q - T_{k,q} - b_q^{(k)}}{\sigma_{q}^{(k)}}\right)\\
    &=\!\frac{1}{2}{P\!H\!M\!I}_\mathrm{HOR} \left(\!1 \!-\! \frac{P_\mathrm{sat, not \: monitored} + P_\mathrm{const, not \: monitored}} {{P\!H\!M\!I}_\mathrm{VERT}+{P\!H\!M\!I}_\mathrm{HOR}}\!\right),
\end{split}
\end{equation}
where the descriptions of the symbols are given in Table \ref{tab:NotationsARAIM}.
Detailed information and mathematical formulations of the ARAIM user algorithm are discussed in \cite{Blanch15:Baseline}.

If the solution separation threshold test does not pass (i.e., a fault is detected), fault exclusion should be attempted. After the exclusion of faulty signals, the HPL should be calculated considering the probability of wrong exclusion. The HPL equation in this case has an additional factor to (\ref{eq:HPL}). Detailed discussions are given in \cite{Blanch15:Baseline}.

\subsection{Prediction of Conservative HPL in Urban Environments}
\begin{figure}
    \centering
    \includegraphics[width=8.5cm]{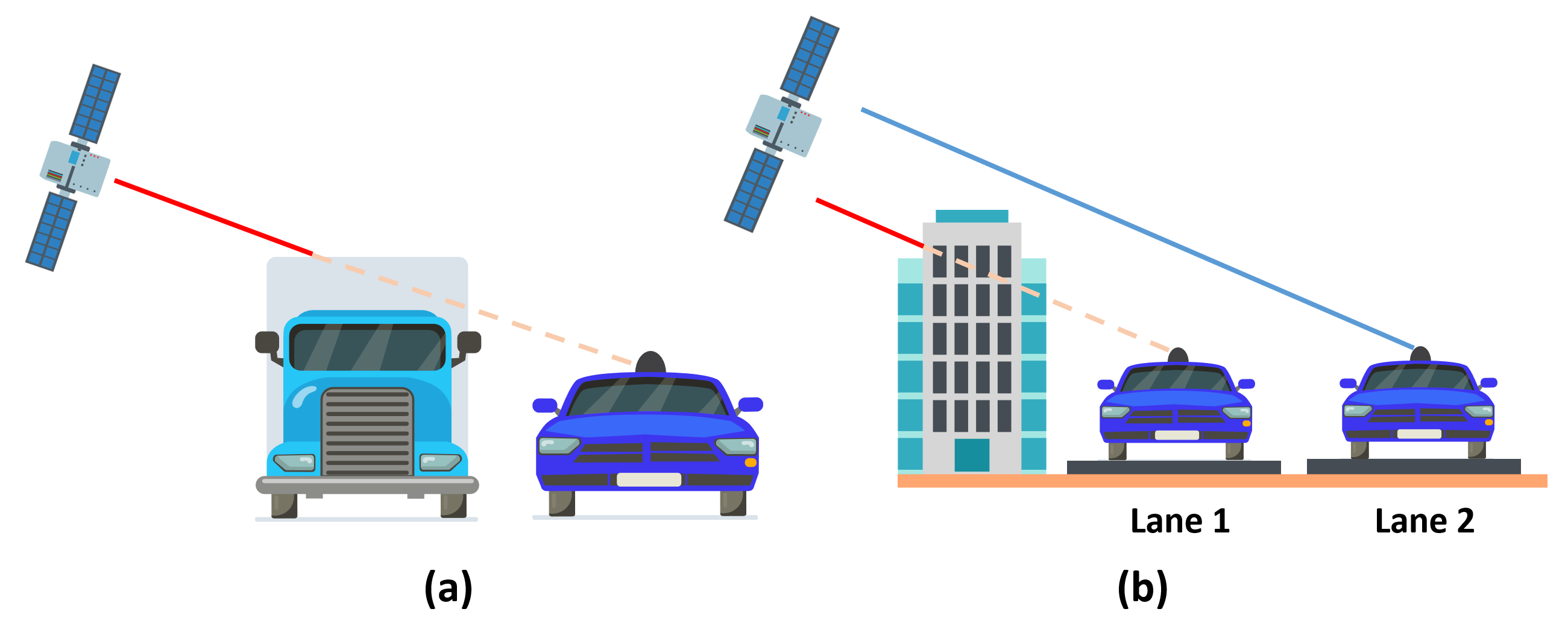}
    \caption{(a) GNSS signal blockage due to a nearby vehicle. (b) Different signal reception conditions at different lanes.}
    \label{fig:truck_and_lane}
\end{figure}

Predicting the \textit{exact} HPL of a vehicle at a certain location and time is virtually impossible due to imperfections in 3D urban digital maps as well as the presence of nearby dynamic objects, which cannot be predicted. For example, nearby vehicles can block satellite signals, as illustrated in Fig. \ref{fig:truck_and_lane}(a). Therefore, the HPL will be predicted conservatively by assuming that the vehicle of interest is always surrounded by taller vehicles. Considering the height of the vehicle used for the field test (1.7 m), the height and width of a typical dump truck (3.3 m and 2.5 m, respectively), and the typical width of a lane (3.7 m), an elevation mask of 33$^\circ$ was set, including a slight margin. In other words, to be conservative, satellite signals with less than 33$^\circ$ elevation are assumed to be blocked by nearby vehicles.

Signal reflection and blockage due to static objects, such as buildings, can be predicted by ray-tracing simulation if the exact 3D urban digital maps are available \cite{Ziedan17:Urban,Zhang18:GNSS}. However, it should be noted that the signal reception conditions at each lane can vary significantly \cite{Miura15:GPS}. For example, a vehicle can have an LOS reception of a certain satellite signal in one lane but may not receive the signal from the same satellite in another lane, because of building blockage (see Fig. \ref{fig:truck_and_lane}(b)).

\begin{figure}
    \centering
    \includegraphics[width=1.0\linewidth]{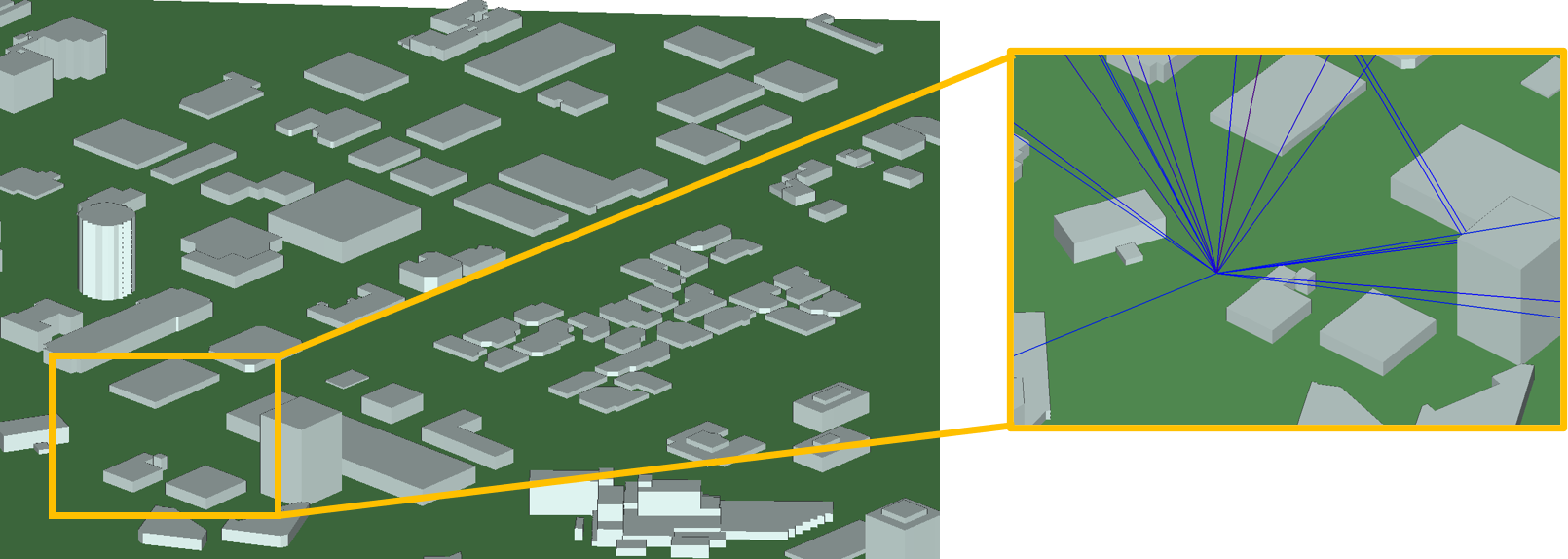}
    \caption{Ray-tracing example at a single node within a 3D urban digital map.}
    \label{fig:ray-tracing}
\end{figure}

To perform ray-tracing simulations to predict signal blockage due to buildings and the NLOS-only or LOS+NLOS bias (i.e., $\rho_\mathrm{NLOS}^{n}$ or $\rho_\mathrm{L+N}^{n}$ in Fig. \ref{fig:multipath_and_NLOS}), commercial 3D urban digital maps from 3dbuildings and Wireless InSite commercial ray-tracing software were used. Fig. \ref{fig:ray-tracing} shows an example of a ray-tracing simulation. It was assumed that the exterior walls of all buildings were made of concrete.
The time-of-arrival (TOA) of GNSS signals was calculated using the shooting and bouncing ray (SBR) method described in \cite{Schuster96:Comparison}, which is used to find geometrical propagation paths between a transmitter and a receiver using a 3D map. In the SBR method, among the rays transmitted from the source, the rays that hit the building are specularly reflected and traced until the maximum number of reflections is reached. Then, $\rho_\mathrm{NLOS}^{n}$ or $\rho_\mathrm{L+N}^{n}$ was predicted using the simulated TOAs, amplitudes, and phases of GNSS signals from ray-tracing according to the signal reception condition. The GPS and GLONASS constellations were considered based on their almanac information.

To reduce the computational complexity of the ray-tracing simulation, it was assumed that the receiver receives only the direct and single reflected signals. If a signal is reflected by buildings more than once, it was assumed that the signal was not received by the vehicle. This assumption does not significantly affect the accuracy of conservative HPL prediction because the received signal strength of multiple reflected signal is low and a receiver may not track such signals.

\begin{figure}
    \centering
    \includegraphics[width=8.5cm]{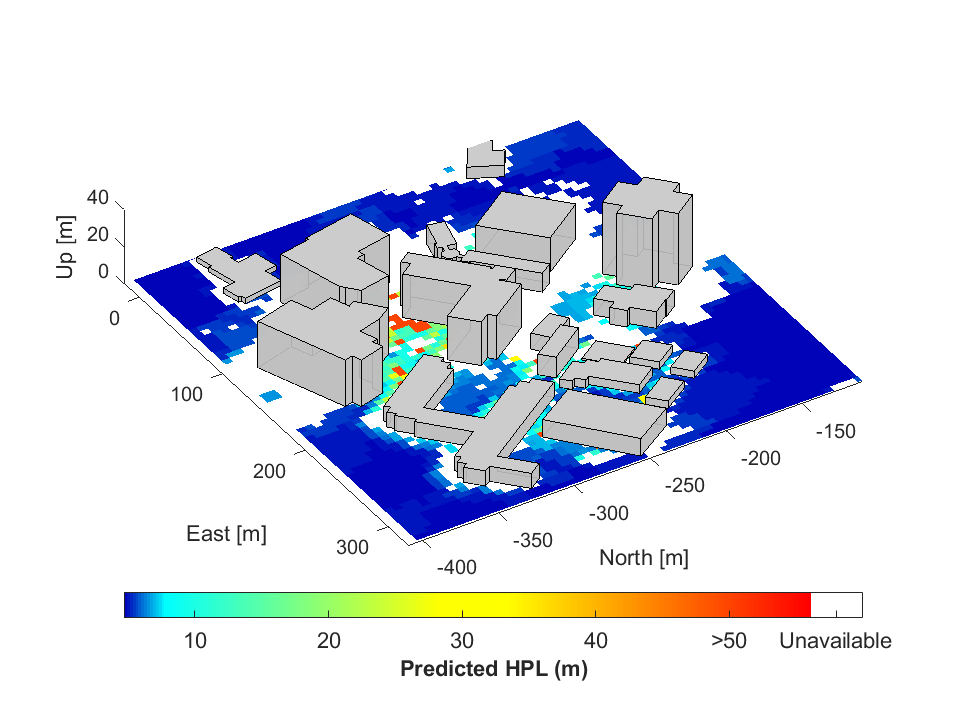}
    \caption{Example map of conservatively predicted HPL with a 33$^\circ$ elevation mask at a certain time epoch. This map varies with time because of GNSS satellite motion.}
    \label{fig:HPL_map}
\end{figure}

With the predicted pseudoranges from the ray-tracing simulation, the HPL can be predicted following the procedure in Fig. \ref{fig:flowchart}. An example map of the conservatively predicted HPL is shown in Fig. \ref{fig:HPL_map}. If the number of visible satellites at a certain location is insufficient for the FDE, the location is marked as unavailable because the HPL prediction is not performed in this case.
It should be noted that the HPL at a given location varies with time because GNSS satellites move. Fortunately, future satellite positions are reliably predictable based on ephemerides \cite{Misra10:Global}. Thus, the conservative HPLs over a certain time horizon at each location can be calculated in advance in a cloud server. Automated vehicles can use this information without worrying about their on-board computational power. Since the conservative HPL prediction at each location and time can be  performed independently, a cloud server with enough parallel processors can quickly generate the HPL prediction maps of the regions of interest.

\subsection{Experimental Field Test Results}
\label{sec:FieldTest}
To verify the proposed methodology for conservatively predicting HPL in urban environments, field tests were performed to calculate the HPL based on actual pseudorange measurements of the experiment. Then, the HPL based on measured pseudoranges (i.e., \textit{measured HPL}) was compared with the conservative HPL based on predicted pseudoranges (i.e., \textit{conservatively predicted HPL}).

HPL varies over time as satellite geometry changes. Further, HPL is impacted by the surrounding environment. To check if the proposed methodology is applicable to various times and environments, field tests were performed in two different cities: Irvine and Riverside, California, USA.

\begin{figure}
    \centering
    \includegraphics[width=8.0cm]{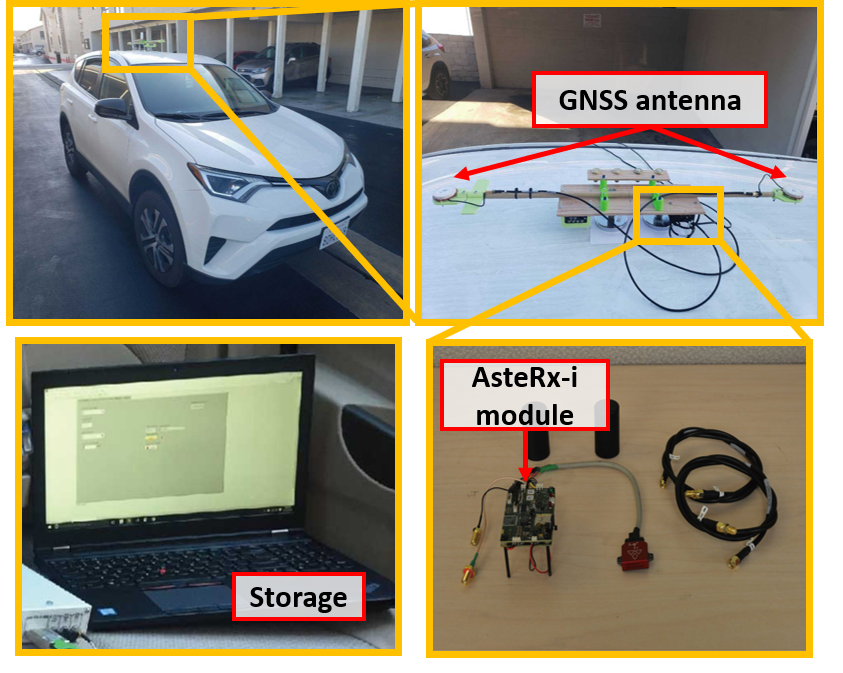}
    \caption{Experimental settings. A GNSS antenna was attached to the top of a ground vehicle. The GNSS signals were processed using a Septentrio's AsteRx-i receiver module. GNSS measurements and navigation data were stored on a laptop placed inside the vehicle.}
    \label{fig:setting}
\end{figure}

\begin{figure}
    \centering
    \includegraphics[width=8.0cm]{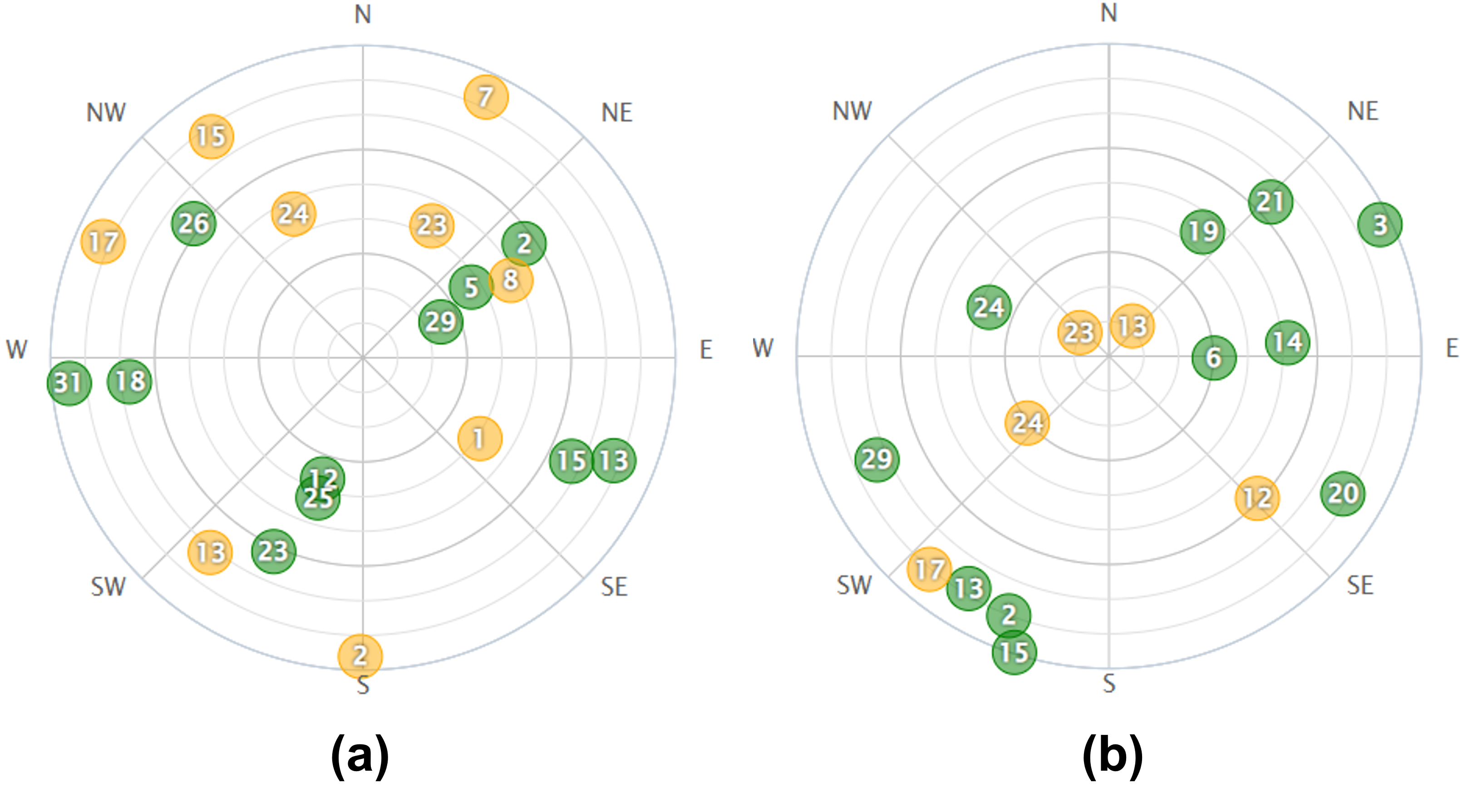}
    \caption{GPS and GLONASS constellations during the field tests in (a) Irvine and (b) Riverside.}
    \label{fig:constellation}
\end{figure}

\begin{figure}
    \centering
    \includegraphics[width=8.0cm]{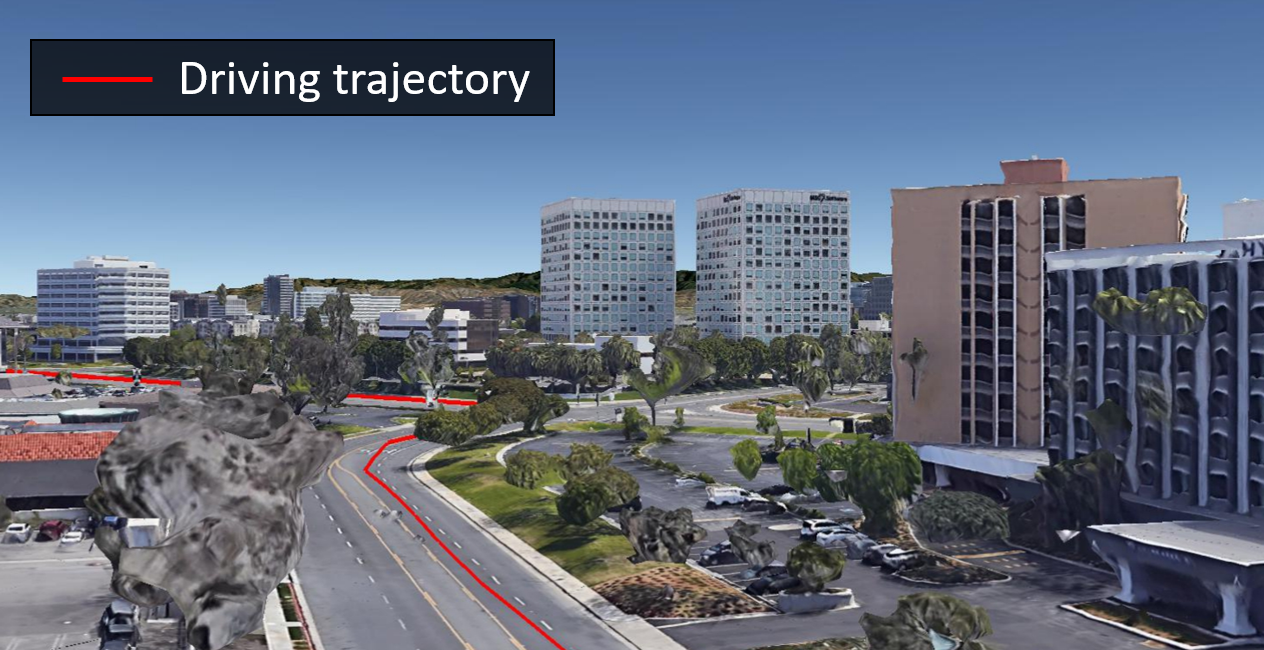}
    \caption{Urban test environment in Irvine. The red line represents the trajectory of the ground vehicle.}
    \label{fig:trajectory_irvine}
\end{figure}

During the experiments, GPS and GLONASS measurements were collected using a Septentrio AsteRx-i V\textsuperscript{\textregistered} receiver. The GNSS antenna was placed on top of the ground vehicle (Fig. \ref{fig:setting}). GNSS constellations during the experiments in Irvine and Riverside are shown in Fig. \ref{fig:constellation}.
Fig. \ref{fig:trajectory_irvine} presents a small portion of the urban test environment in Irvine as an example, which comprised several tall buildings that  significantly changed the measured HPL values.
In Riverside, complex-shaped buildings were distributed along the test trajectory.
The experiments were conducted along approximately 4.5 km and 1.6 km roads in Irvine and Riverside, respectively.

As shown in Fig. \ref{fig:truck_and_lane}(b), the signal reception condition can dramatically change according to the lateral location of a vehicle on the road. It is theoretically possible to predict the HPL at every location, as shown in Fig. \ref{fig:HPL_map}; however, the prediction accuracy depends on the accuracy of the 3D building and road maps. For example, a slight height error of a building model or a lateral position error of a road model in a digital map can cause a visible satellite to be predicted as invisible during the ray-tracing simulation. Unfortunately, commercially available 3D digital maps have limited accuracy. As a conservative approach, multiple ray-tracing simulations were performed by changing the vehicle's lateral location across the road. If a certain satellite is invisible at one location, the satellite is treated as an invisible satellite when predicting the HPL at the given longitudinal location of the road. Furthermore, $\rho_\mathrm{NLOS}^{n}$ and $\rho_\mathrm{L+N}^{n}$ were also predicted at every lateral location across the road, and the largest value was chosen for the pseudorange prediction, to be conservative.

\begin{figure}
    \centering
    \includegraphics[width=7.5cm]{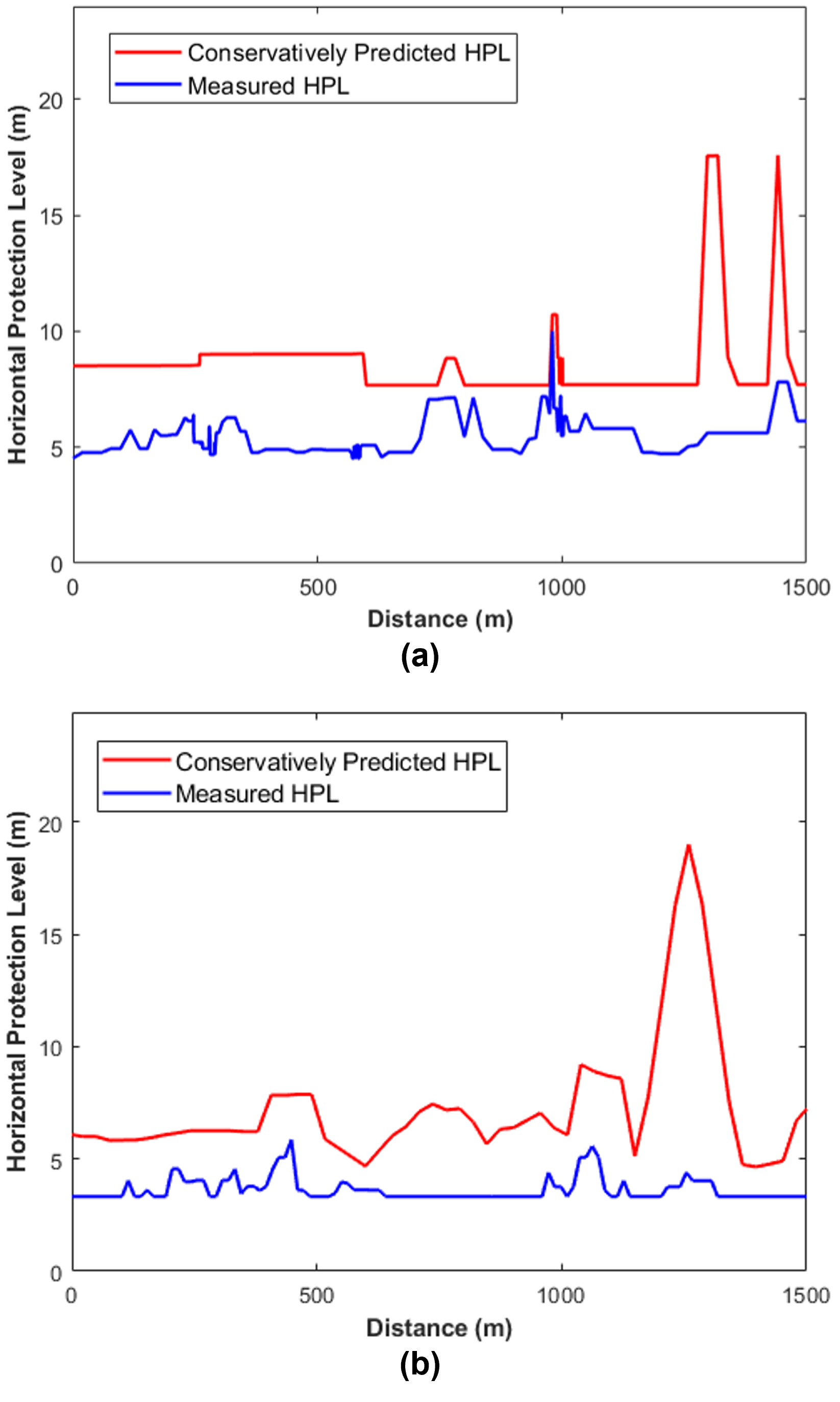}
    \caption{Experimental results in (a) Irvine and (b) Riverside. Conservatively predicted HPLs overbounded measured HPLs along sample paths.}
    \label{fig:HPL_graph}
\end{figure}

Fig. \ref{fig:HPL_graph} shows the conservatively predicted HPL along two 1.5-km roads with tall buildings. The ground vehicle freely changed its driving lane during the field tests. However, its measured HPL was always less than the conservatively predicted HPL that assumed the most challenging lateral location, having the largest number of signal blockages and largest NLOS-only and LOS+NLOS biases. When the vehicle drove along a lane with a better satellite visibility (i.e., a distant lane from a tall building), the measured HPL was significantly lower than the conservatively predicted HPL that assumed the most challenging lane with a poor satellite visibility, as in the case of a 1.3-km distance location in Fig. \ref{fig:HPL_graph}(b). Nevertheless, the most challenging lane needs to be assumed when HPL is predicted because it is not practical to restrict the driving lane of a vehicle.

\section{Application Case Study: Safety-Constrained Path Planning}
\label{Application_Case_Study_Safety_Constrained_Path_Planning}

The predicted satellite navigation reliability map (i.e., HPL prediction map) can be utilized by an automated vehicle for various purposes to ensure safe driving. Because the reliability of satellite navigation signals is already known through the HPL prediction map, an automated vehicle can plan a safe trajectory ahead of time. If the navigation sensors of the vehicle rely heavily on GNSS, it would be better to detour around the high HPL region. Most automated vehicles utilize IMUs, which are calibrated using GNSS. Therefore, the IMU outputs in the high HPL region should not be relied upon.

As an application case study, path planning of an automated vehicle based on the HPL prediction map is considered. Unlike traditional strategies for path planning to minimize travel distance and time, the primary focus here is the navigation safety of an automated vehicle. Therefore, the optimization problem is formulated with safety considerations as
\begin{align} \label{eq:path_optimization}
\begin{split}
    \underset{\pi \in \mathcal{P}}{\text{minimize}} & \quad
    \sum_{p_k \in \pi} {dist}(p_{k-1}, p_k) \cdot  {H\!P\!L}(p_k,t)\\
    \text{subject to}
    & \quad
    \frac{N\!\left({H\!P\!L}(p_{k},t) < T_\mathrm{HPL}\right)}{N_\mathrm{nodes}} > T_\mathrm{safe} \\
    &\quad
    D_\mathrm{HPL \: unacceptable} < D_\mathrm{safe},
\end{split}
\end{align}
where the descriptions of the symbols are given in Table \ref{tab:NotationsPathPlanning}.

\begin{table}
\centering
\caption{Mathematical Notations Related to the Proposed Safety-Constrained Path Planning Algorithm.}
\label{tab:NotationsPathPlanning}
\vspace{-6mm}
\begin{center}
{\renewcommand{\arraystretch}{1.4}
 \begin{tabular}[cl]{>{\centering\arraybackslash}m{2.2cm}|>{\arraybackslash}m{5.8cm}}
 \Xhline{2\arrayrulewidth}
    \thead{Symbol} & \thead{Description} \\
\Xhline{1.5\arrayrulewidth}
{$\pi$} & {Sequence of nodes between the start node $p_\mathrm{start}$ and target node $p_\mathrm{target}$,
i.e., $\pi = \{p_{\mathrm{start}}, p_2, p_3, \cdots , p_{\mathrm{target}} \}$} \\
{$N_\mathrm{nodes}$} & {Total number of nodes along a path} \\
{$dist\left(p_{k-1}, p_{k}\right)$} & {Euclidean distance between nodes $p_{k-1}$ and $p_{k}$ ($p_1 = p_\mathrm{start}$ and $p_{N_\mathrm{nodes}} = p_\mathrm{target}$)} \\
{$H\!P\!L\left(p_{k}, t\right)$} & {Conservatively predicted HPL at node $p_{k}$ and time $t$, which is given by the HPL prediction map} \\
{$T_\mathrm{HPL}$} & {Maximum allowable HPL value (i.e., HPL threshold)} \\
{$N\left(\cdot\right)$} & {Number of nodes satisfying the given condition} \\
{$T_\mathrm{safe}$} & {Threshold for the ratio of nodes satisfying the HPL threshold $T_\mathrm{HPL}$} \\
{$D_\mathrm{HPL \: unacceptable}$} & {Continuous distance where the predicted HPL is unavailable or above $T_\mathrm{HPL}$} \\
{$D_\mathrm{safe}$} & {Threshold for the $D_\mathrm{HPL \: unacceptable}$} \\
 \Xhline{2\arrayrulewidth}
\end{tabular}}
\end{center}
\end{table}


The cost function in (\ref{eq:path_optimization}) aims to find an optimal path that minimizes both the travel distance and HPL along the path (recall that a smaller HPL indicates a higher satellite navigation reliability). The first constraint in (\ref{eq:path_optimization}) considers the ratio of the number of safe nodes to that of total nodes. For example, if $T_\mathrm{HPL}$ is set to 10 m and $T_\mathrm{safe}$ is set to 95\%, a candidate path with more than 5\% of nodes having an HPL of over 10 m will not be selected as an optimal path. The second constraint in (\ref{eq:path_optimization}) ensures the avoidance of a candidate path with continuous signal outages. The outputs from the automotive-grade IMUs quickly diverge if the GNSS signals are unavailable or unreliable for a certain period. Therefore, continuous signal outages are more problematic than intermittent signal outages for similar total outage duration. For example, if $D_\mathrm{safe}$ is set to 150 m, a candidate path with continuous signal outages for more than 150 m distance will not be selected as an optimal path.

Table \ref{tab:Comparison_Metrics} compares the optimization problem formulations of the previous studies \cite{Zhang19:New,Maaref20:Optimal} and the current study. Unlike the previous studies, where only travel distance and navigation reliability (i.e., positioning error \cite{Zhang19:New} or HPL without considering measurement faults \cite{Maaref20:Optimal}) were considered, the proposed optimization problem considers the GNSS unavailability and continuous signal outages as well to obtain a more realistic solution.

\begin{table} [b]
\centering
\caption{Comparison of Optimization Problem Formulations for Safety-Constrained Path Planning.}
\label{tab:Comparison_Metrics}
\vspace{-4mm}
\begin{center}
{\renewcommand{\arraystretch}{1.6}
 \begin{tabular}[c]{>{\centering\arraybackslash}m{1.35cm}|>{\centering\arraybackslash}m{1cm}>{\centering\arraybackslash}m{1.25cm} >{\centering\arraybackslash}m{1.5cm}>{\centering\arraybackslash}m{1.4cm}}
 \Xhline{2\arrayrulewidth}
    \vspace{2pt} \thead{Method} & \vspace{-2pt}\thead{Travel \\ distance}
       & \vspace{-2pt}\thead{Navigation \\ reliability} & \vspace{-2pt}\thead{GNSS \\ unavailability} & \vspace{-10pt}\thead{Continuous \\ GNSS \\ outage}\\
\Xhline{1.5\arrayrulewidth}
 {Zhang and Hsu \cite{Zhang19:New}} & \checkmark & \checkmark & \xmark & \xmark \\
 \hline
 {Maaref and Kassas \cite{Maaref20:Optimal}} & \checkmark & \checkmark & \xmark & \xmark \\
  \hline
 {Proposed} & \checkmark & \checkmark & \checkmark &\checkmark \\
 \Xhline{2\arrayrulewidth}
\end{tabular}}
\end{center}
\end{table}

To solve the optimization problem in (\ref{eq:path_optimization}), the A* algorithm \cite{Russell10:Artificial} was applied, which is a widely-used search algorithm that can find an optimal path to a given target node.
The A* algorithm was implemented as shown in Algorithm \ref{alg:PathPlanning} to find an optimal solution of the safety-constrained path planning problem.
The overall road structure of a given map, which is expressed by a graph composed of nodes and edges, is denoted by $\mathcal{P}$. Given start and target nodes, the A* algorithm finds the cheapest path (i.e., a sequence of nodes that minimizes the cost function in (\ref{eq:path_optimization})) based on the sum of  backward cost (cumulative cost) and forward cost (heuristic cost).
The open set, which is implemented as a priority queue that stores the nodes that have been visited but their successors are not explored, is denoted by $\mathcal{O}$. $p_{\mathrm{current}}$ denotes the currently visited node, and $p_{\mathrm{neighbor}}$ denotes a neighbor node of $p_{\mathrm{current}}$.
For each iteration, all neighbor nodes of $p_{\mathrm{current}}$ are stored in $\mathcal{O}$ and the overall cost $f$ of each neighbor node is calculated.
The overall cost $f$ is defined as the sum of  cumulative cost $g$ and heuristic cost $h$. The Euclidean distance (i.e., straight-line distance) to the target node was used as the heuristic cost.
After calculating the cost of each neighbor node, the node in $\mathcal{O}$ with the smallest $f$ is selected as $p_{\mathrm{current}}$ and is moved to the close set $\mathcal{C}$.
The iteration ends when the target node is reached or when the open set $\mathcal{O}$ becomes empty.
If the target node is reached, the final optimal path $\pi$ can be found by reconstructing the nodes in $C$.

\begin{algorithm}
\caption{A* algorithm implementation for the safety-constrained path planning.}
\label{alg:PathPlanning}
\vspace{0.2em}
\KwData{$\mathcal{P}$, $p_{\mathrm{start}}$, $p_{\mathrm{target}}$, ${H\!P\!L}$, $D_\mathrm{safe}$, $T_\mathrm{HPL}$}
\KwResult{$\pi$}
$f\left(p_{\mathrm{start}}\right) \gets dist\left(p_{\mathrm{start}}, p_{\mathrm{target}}\right)$\\
\vspace{0.1em}
$D_{\mathrm{HPL \: unacceptable}}\left(p_{\mathrm{start}}\right) \gets 0$\\
\vspace{0.2em}
$\mathit{safenode}\left(p_{\mathrm{start}}\right) \gets 1$\\
$\mathcal{O} \gets p_{\mathrm{start}}$\\
\vspace{0.2em}
\While{$\mathcal{O} \text{ \upshape is not empty}$}{
    \vspace{0.2em}
    $p_{\mathrm{current}} \gets $ node in $\mathcal{O}$ having smallest $f$\\

    $\mathcal{O} \gets \mathcal{O} - p_{\mathrm{current}}$\\
    \If{\upshape $D_{\mathrm{HPL \: unacceptable}}\left(p_{\mathrm{current}}\right) \geq D_\mathrm{safe}$} {
        \vspace{0.2em}
        \Continue
        \vspace{0.2em}
    }
    \If{\upshape $p_{\mathrm{current}}$ is $p_{\mathrm{target}}$} {
    \vspace{0.2em}
    $\pi \gets$ reconstructed path from $\mathcal{C}$\\
    $N_\mathrm{nodes} \gets$ total number of nodes in $\pi$\\
    \vspace{0.2em}
    $N_\mathrm{safe\:nodes} \gets$ \newline sum of $\mathit{safenode}$ of all nodes in $\pi$\\
    \vspace{0.2em}
    \If{\upshape ${N_\mathrm{safe\:nodes}} / {N_\mathrm{nodes}} > T_\mathrm{safe}$}{
            \vspace{0.2em}
            \Return $\pi$
            \vspace{0.2em}
    }
    \vspace{0.2em}
    \Continue
    \vspace{0.2em}
    }

    $\mathcal{C} \gets \mathcal{C} + p_{\mathrm{current}}$\\
    \vspace{0.2em}
    \For{\upshape every neighbor of $p_{\mathrm{current}}$}{
        $g\left(p_{\mathrm{neighbor}}\right) \gets
        \newline dist\left(p_{\mathrm{neighbor}}, p_{\mathrm{current}}\right) \cdot {H\!P\!L}\left(p_{\mathrm{neighbor}}\right)
        \newline \hspace*{3em} + g\left(p_{\mathrm{current}}\right)$\\
        $h\left(p_{\mathrm{neighbor}}\right) \gets dist\left(p_{\mathrm{neighbor}}, p_{\mathrm{target}}\right)$\\
        $f\left(p_{\mathrm{neighbor}}\right) \gets g\left(p_{\mathrm{neighbor}}\right) + h\left(p_{\mathrm{neighbor}}\right)$\\
        \vspace{0.2em}

        \eIf{\upshape  ${H\!P\!L}\left(p_{\mathrm{neighbor}}\right)$ is unacceptable} {
            \vspace{0.2em}
            $\mathit{safenode}\left(p_{\mathrm{neighbor}}\right) \gets 0$\\
            $D_{\mathrm{HPL \: unacceptable}}\left(p_{\mathrm{neighbor}}\right) \gets$ \newline $D_{\mathrm{HPL \: unacceptable}}\left(p_{\mathrm{current}}\right) \newline \hspace*{1em} + dist\left(p_{\mathrm{neighbor}}, p_{\mathrm{current}}\right)$\\
        }
        {
            $\mathit{safenode}\left(p_{\mathrm{neighbor}}\right) \gets 1$\\
            $D_{\mathrm{HPL \: unacceptable}}\left(p_{\mathrm{neighbor}}\right) \gets 0$\\
        }
        $\mathcal{O} \gets \mathcal{O} + p_{\mathrm{neighbor}}$\\
        }
}
\Return failure
\vspace{0.2em}
\end{algorithm}

Considering the four candidate paths shown in Fig. \ref{fig:paths} from Costa Mesa to Irvine, California, USA, the key metrics related to the optimization problem in (\ref{eq:path_optimization}) along each candidate path are summarized in Table \ref{tab:path_planning}.
The GPS and GLONASS pseudoranges were measured along the paths during the field tests to obtain the measured HPL. The results of this experiment are summarized as follows:

\begin{figure}
    \centering
    \includegraphics[width=8.6cm]{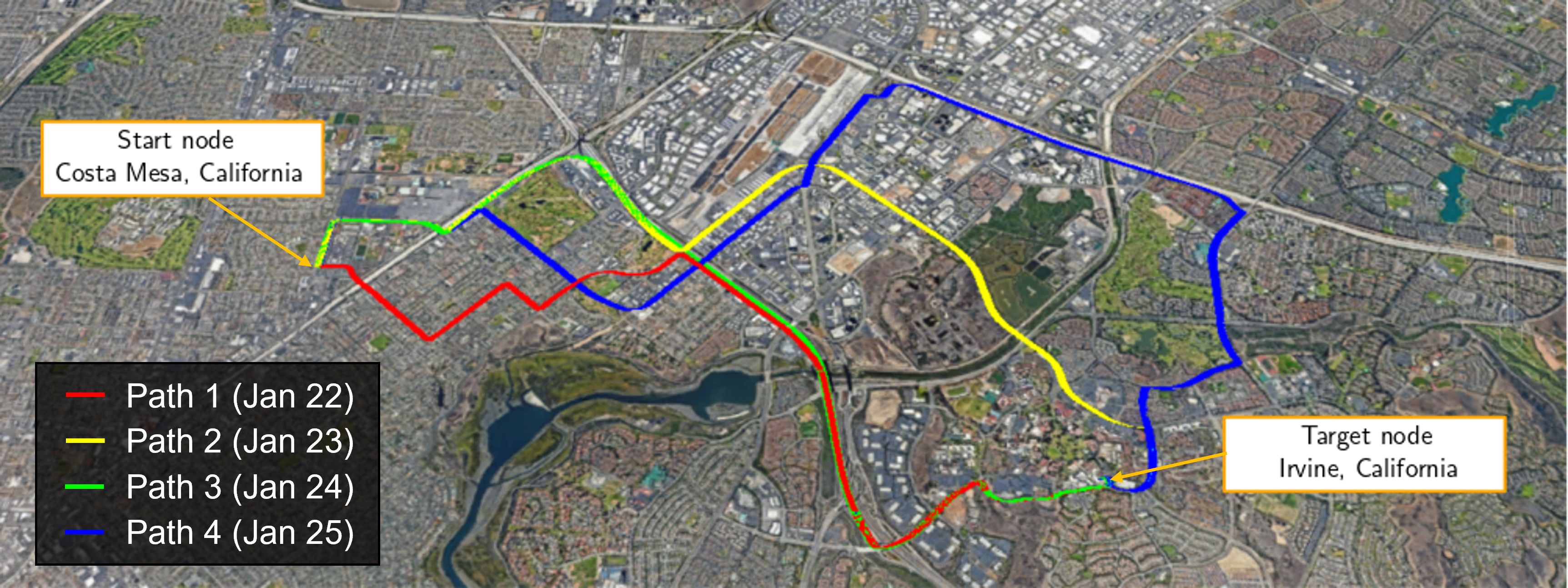}
    \caption{Four candidate paths from Costa Mesa to Irvine. GNSS signals along four paths were collected over four consecutive days.}
    \label{fig:paths}
\end{figure}

\begin{list}{$\bullet$}{\leftmargin=1em \itemindent=0em}
    \item The costs, which are the output of the cost function in (\ref{eq:path_optimization}), of paths 1, 2, 3, and 4 were $56428$, $52137$, $110398$, and $92805$, respectively. Therefore, path 2 has the minimum cost. Because path 2 satisfies all the constraints in (\ref{eq:path_optimization}), it is selected as the optimal path.
    \item Although the average HPLs of the four paths were similar, the ratios of safe nodes and the maximum continuous distances with unacceptable HPLs (i.e., predicted HPL is unavailable or above $T_\mathrm{HPL}$) were significantly different. In particular, in path 2, the ratio of safe nodes was 100\% and there was no section where predicted HPL was unacceptable. This implies that an autonomous vehicle can know path 2 has better GNSS signal quality than the other paths before driving by solving the optimization problem in (\ref{eq:path_optimization}) using the HPL prediction map and Algorithm \ref{alg:PathPlanning}.
    \item Paths 1 and 4 are also feasible solutions because they satisfied all the constraints of (\ref{eq:path_optimization}). However, path 1 or 4 is not an optimal solution according to the proposed cost function that considers both travel distances and predicted HPLs.
    \item Path 3 is not a feasible solution because it violated the second constraint that requires $D_\mathrm{HPL \: unacceptable}$ to be less than $D_\mathrm{safe}$ which was set to 150 m. The proposed optimization problem successfully screened a path with continuous GNSS signal outages that can potentially threaten the vehicle's driving safety.
    \item In all cases, the conservatively predicted HPL bounded the measured HPL 100\% of the time.
\end{list}

\begin{table}
\centering
\caption{Comparison of Key Optimization Metrics along Four Candidate Paths.}\label{tab:path_planning}
 \begin{tabular}{ c | c | c | c | c | c }
 \hhline{======}
 \thead{Path} & \thead{Travel \\ distance \\ {[m]}} & \thead{Average \\ predicted \\ HPL \\ {[m]}} & \thead{Average \\ measured \\ HPL \\ {[m]}} & \thead{Ratio of \\ safe \\ nodes \\ {[\%]}} & \thead{Maximum \\ continuous \\ distance \\ with \\ unacceptable \\ HPL [m]}
\\
 \hline
 Path 1 & 9,746 & 6.49 & 5.57 & 98.5 & 131.90 \\
 \hline
 Path 2 & 9,631 & 7.91 & 5.60 & 100 & 0 \\
 \hline
 Path 3 & 14,244 & 7.67 & 5.52 & 97.2 & 208.72  \\
 \hline
 Path 4 & 10,629 & 8.50 & 5.64 & 97.1 & 103.95 \\
 \hhline{======}
\end{tabular}
\end{table}

\section{Conclusion}
\label{Conclusion}
The reliability of GNSS signals is crucial to ensure driving safety, because various navigation sensors of automated vehicles rely on GNSS signals. This article considered the HPL obtained by the ARAIM algorithm as a metric to measure the navigation reliability at a given location and time on urban roads.
Due to the uncertainty of nearby dynamic objects and the limited accuracy of 3D urban digital maps, a method to conservatively predict the HPL was proposed and validated experimentally.
The pseudorange biases and the presence of signal reflections and blockages, which are necessary to predict HPL, in urban environment were simulated by ray-tracing with 3D maps.
The generated HPL prediction map can serve as useful road information for various navigation applications.
As a case study, the HPL prediction map was applied for safety-constrained path planning of an automated ground vehicle.
Unlike the previous studies, the proposed optimization problem considered the unavailability of GNSS signals and continuous GNSS signal outages that occur in urban environments. A specific implementation of the A* algorithm to find an optimal path was also suggested and demonstrated.


\section*{Acknowledgment}
The authors would like to thank Mahdi Maaref for data collection and insightful discussions. This work was supported in part by the Ministry of Science and ICT (MSIT), Korea, under the High-Potential Individuals Global Training Program (2020-0-01531) supervised by the Institute for Information \& Communications Technology Planning \& Evaluation (IITP), in part by the Unmanned Vehicles Core Technology Research and Development Program through the National Research Foundation of Korea (NRF) and the Unmanned Vehicle Advanced Research Center (UVARC) funded by the Ministry of Science and ICT, Republic of Korea (2020M3C1C1A01086407), and in part by the Basic Science Research Program through the National Research Foundation of Korea (NRF) funded by the Ministry of Education (2021R1A6A3A13046688).
This work was supported in part by the National Science Foundation (NSF) under Grant 1929965 and in part by the U.S. Department of Transportation (USDOT) under Grant 69A3552047138 for the CARMEN University Transportation Center (UTC).

\bibliographystyle{IEEEtran}
\bibliography{references}

%

\vspace{-0.85cm}
\begin{IEEEbiography}[{\includegraphics[width=1in,height=1.25in,clip,keepaspectratio]{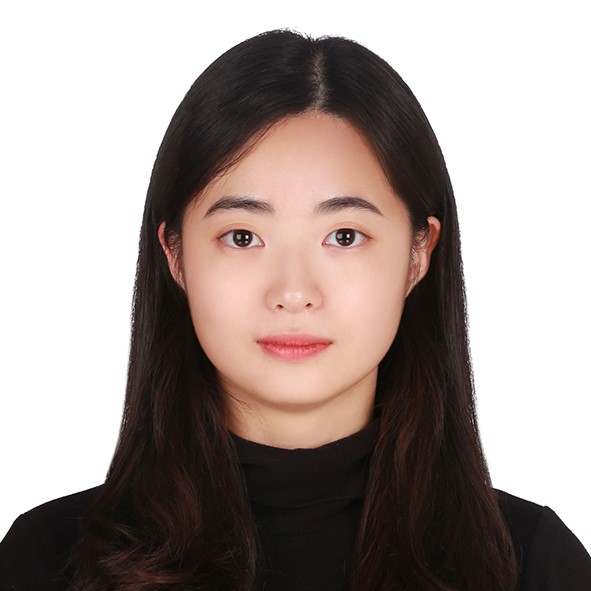}}]{Halim Lee}
(halim.lee@yonsei.ac.kr) is a Ph.D. student in the School of Integrated Technology, Yonsei University, Incheon, Korea. She received the B.S. degree in Integrated Technology from Yonsei University. She was a visiting graduate student with the Autonomous Systems Perception, Intelligent, and Navigation (ASPIN) Laboratory at the University of California, Irvine. Her research interests include motion planning, integrity monitoring, and opportunistic navigation.
\end{IEEEbiography}

\vspace{-0.7cm}
\begin{IEEEbiography}[{\includegraphics[width=1in,height=1.25in,clip,keepaspectratio]{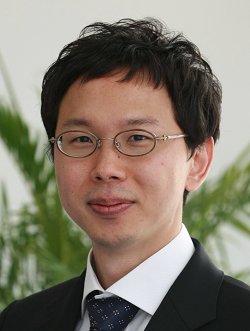}}]{Jiwon Seo}
(jiwon.seo@yonsei.ac.kr) is an associate professor with the School of Integrated Technology, Yonsei University, Incheon, Korea. He received the B.S. degree in mechanical engineering (division of aerospace engineering) in 2002 from Korea Advanced Institute of Science and Technology, Daejeon, Korea, and the M.S. degree in aeronautics and astronautics in 2004, the M.S. degree in electrical engineering in 2008, and the Ph.D. degree in aeronautics and astronautics in 2010 from Stanford University, Stanford, CA, USA.
His research interests include GNSS and complementary PNT systems. Prof. Seo is a member of the International Advisory Council of the Resilient Navigation and Timing Foundation, Alexandria, VA, USA, and a member of several advisory committees of the Ministry of Oceans and Fisheries and the Ministry of Land, Infrastructure and Transport, Korea.
\end{IEEEbiography}

\vspace{-0.5cm}
\begin{IEEEbiography}[{\includegraphics[width=1in,height=1.25in,clip,keepaspectratio]{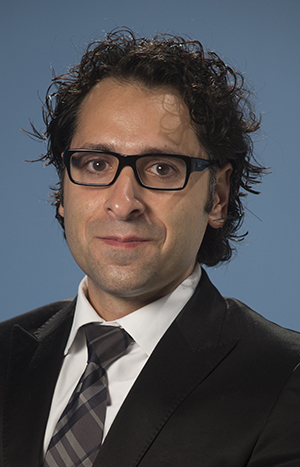}}]{Zaher M. Kassas}
(zkassas@ieee.org) is an associate professor at the University of California, Irvine and director of the Autonomous Systems Perception, Intelligence, and Navigation (ASPIN) Laboratory. He is also director of the U.S. Department of Transportation Center: CARMEN (\underline{C}enter for \underline{A}utomated Vehicle \underline{R}esearch with \underline{M}ultimodal Assur\underline{E}d \underline{N}avigation), focusing on navigation resiliency and security of highly automated transportation systems. He received a B.E. in Electrical Engineering from the Lebanese American University, an M.S. in Electrical and Computer Engineering from The Ohio State University, and an M.S.E. in Aerospace Engineering and a Ph.D. in Electrical and Computer Engineering from The University of Texas at Austin. He is a recipient of the 2018 National Science Foundation (NSF) Faculty Early Career Development Program (CAREER) award, 2019 Office of Naval Research (ONR) Young Investigator Program (YIP) award, 2022 Air Force Office of Scientific Research (AFOSR) YIP award, 2018 IEEE Walter Fried Award, 2018 Institute of Navigation (ION) Samuel Burka Award, and 2019 ION Col. Thomas Thurlow Award. He is a Senior Editor of the IEEE Transactions on Intelligent Vehicles and an Associate Editor of the IEEE Transactions on Aerospace and Electronic Systems and the IEEE Transactions on Intelligent Transportation Systems. His research interests include cyber-physical systems, estimation theory, navigation systems, autonomous vehicles, and intelligent transportation systems.
\end{IEEEbiography}

\vfill

\end{document}